\documentclass[12pt, draftclsnofoot, onecolumn]{IEEEtran}
\usepackage{filecontents}
\usepackage{amsthm}
\usepackage[table,xcdraw]{xcolor}
\newtheorem{theorem}{\textbf{\text{Theorem}}}

\usepackage{epsfig}
\usepackage{times}
\usepackage{verbatim}
\usepackage{enumerate}
\usepackage{multicol}
\usepackage{afterpage}
\usepackage{wrapfig}
\usepackage{cite}
\usepackage{graphicx}
\usepackage[caption=false]{algorithm}
\usepackage{algorithmic}
\usepackage{ifmtarg}
\usepackage{amssymb}
\usepackage{amsmath}
\usepackage{color}
\usepackage{bbm}
\usepackage{epstopdf}
\usepackage{flushend}
\usepackage{subcaption}

 \usepackage{multirow}
 \usepackage{stackengine}

\makeatletter
\newcommand*{\rom}[1]{\expandafter\@slowromancap\romannumeral #1@}

\makeatother
\begin{document}
\title{Escaping the Densification Plateau in Cellular Networks Through mmWave Beamforming}
	\author{
		\IEEEauthorblockN{\large  Ahmad AlAmmouri,  Manan Gupta,  Fran\c cois Baccelli, and Jeffrey G. Andrews }\\
		\thanks{The authors are with the Wireless Networking and Communications Group (WNCG), The University of Texas at Austin, Austin, TX 78712 USA. Email: \{alammouri@utexas.edu, g.manan@utexas.edu, francois.baccelli@austin.utexas.edu, jandrews@ece.utexas.edu\}. Part of this material is based upon research supported by the Chateaubriand Fellowship of the Office for Science $\&$ Technology of the Embassy of France in the United States.}}
	
	\maketitle
	\begin{abstract}
        We study how dense multi-antenna millimeter wave (mmWave) cellular network performance scales in terms of the base station (BS) spatial density $\lambda$, by studying the signal-to-interference-plus-noise ratio (SINR) and the area spectral efficiency (ASE).  If the number of antennas at each BS scales at least linearly with $\lambda$, which increases the number of possible beam configurations and their main-lobe gain, and decreases their side-lobe gain, we prove that the SINR approaches a finite random variable that is independent of $\lambda$ and the ASE scales at least linearly with $\lambda$. In contrast, if the  number of antennas scales sub-linearly with $\lambda$, then the SINR decays to zero and the ASE saturates to a constant.  Thus, by moving to higher carrier frequencies with successively smaller antennas, and exploiting the correspondingly increased directionality, cellular operators can in principle avoid the densification plateau (or collapse) in cellular networks and instead continue to harvest linear sum throughput gains through BS densification.
	\end{abstract}
	%\begin{IEEEkeywords}
%Area spectral efficiency, cellular networks, MIMO, mmWave, stochastic geometry, ultradense networks.
%\end{IEEEkeywords}

\section{Introduction}\label{Sec:Intro}
Network densification has long provided cellular operators with a straightforward way to increase the spatial and per user throughput of their networks, and has been the key driver in throughput gains over several cellular network generations.  However, the sustainability of densification for 5G and beyond has been called into question \cite{Are_Andrews16}, with the main argument being that the linear growth observed so far in practice and supported by earlier mathematical models \cite{A_Andrews11} does not survive under more realistic path loss models and analyses that better describe short communication distances and dense networks \cite{Performance_Ding17,Performance_Nguyen17,SINR_AlAmmouri17,The_Renzo16}.

Recently in \cite{A_AlAmmouri19}, we unified the previous results and showed that under natural assumptions on the network and the signal propagation models, a densification plateau, where densifying the network sacrifices the per-user throughput and yields diminishing gains in terms of the ASE, is inevitable. Nevertheless, \cite{A_AlAmmouri19} focused on single-antenna BSs and users' equipment (UEs) and thus omnidirectional transmission.  More recently, we showed in \cite{Area_AlAmmouri20} that by deploying multi-antenna BSs, with a number of antennas that scales at least linearly with the BS spatial density, we could ensure a non-zero per-user throughput and obtain a linear gain of ASE with densification. Hence, the densification plateau can be avoided in theory, echoing an argument that was motivated by earlier works on MIMO ad hoc networks \cite{Spectral_Lee16}. 

However, the works in \cite{Area_AlAmmouri20,Spectral_Lee16} focused on traditional cellular frequency bands, i.e., sub-6 GHz, where the channel seen by different transmit and receive antennas are assumed to be independent and identically distributed (i.i.d.). Hence, the practicality of the results can be questioned since by scaling the number of antennas with the BS  spatial density, the size of antenna arrays grows and very soon becomes infeasible, especially in the context of the small cells. To this end, millimeter wave (mmWave) communications is an attractive technology due to the directionality of the antenna arrays, and the small size of these arrays \cite{Millimeter_Rappaport13}. However, one cannot directly generalize the results from \cite{Area_AlAmmouri20} and claim that we can harvest linear ASE gains with network densification, since the signal propagation is very different at these bands, and the i.i.d. assumption, alluded to above, is essential for the results in \cite{Area_AlAmmouri20} to hold. Hence, the objective of this work is to answer the question of whether or not we can achieve similar performance gains as in \cite{Area_AlAmmouri20} for mmWave bands.

Note that the scaling laws of mmWave networks have been studied in \cite{Performance_Thornburg16,Ergodic_Thornburg18} for ad hoc networks. In \cite{Performance_Thornburg16}, the authors focused on the coverage probability, and in \cite{Ergodic_Thornburg18} they focused on the ASE, which makes it more relevant to this work. The key result in \cite{Ergodic_Thornburg18} is that we can ensure a non-zero per-user throughput and a linear ASE gain with densification by scaling the number of antennas with the nodes' spatial density. However, the scaling results in \cite{Ergodic_Thornburg18} were derived under certain assumptions: ($i$) ad hoc network, ($ii$) the power-law path loss model without any consideration for blockages, which are critical in mmWave communications, and ($iii$) a specific distribution for the small scale fading. 

In this work, we relax these specific assumptions by assuming a cellular network, a general physically feasible path loss model which can capture the blockage effects, and a general small scaling fading distribution that depends on the nature of the communication link (LoS/NLoS).  Under this relaxed model, we derive the scaling laws of the SINR and the ASE and prove that scaling the number of antennas at the BSs linearly with the BS spatial density, {which increases the number of beams, increases their main-lobe gain, and decreases their side-lobe gain,} is sufficient to maintain a non-zero per-user throughput and a linear ASE gain. If the number of antennas is scaled sub-linearly with the density, then we show through our simulations that the SINR decays to zero and the ASE saturates to a constant.  

Overall, our results show that the scaling laws derived for traditional cellular frequency bands in \cite{Area_AlAmmouri20} extend to mmWave cellular networks as well. From another perspective, our results show that given a fixed antenna array area, the carrier frequency has to increase as $\sqrt{\lambda}$ (assuming a 2D planar array) to avoid the densification plateau.  To provide a concrete example, if we wish to increase the BS density of an already dense mmWave network at a carrier frequency $f_c = 28$ GHz by a factor of $10$ and still achieve a 10x increase in ASE without increasing the antenna array's physical area, then we would need to increase $f_c$ by a factor of $\sqrt{10}$, i.e. to about 90 GHz, along with the correspondingly larger number of antennas and beamforming gain.  Note that we do not include the beam training overheads in our analysis, which is typical in studying scaling laws in mmWave networks \cite{Performance_Thornburg16,Ergodic_Thornburg18}, just as the channel estimation overheads are not typically included in studying the scaling laws in traditional MIMO networks. Hence, the scaling laws we derive are upper-bounds on the actual network performance we observe in practice.

\section{System Model}\label{Sec:SysMod}
We consider a cellular network, where the BSs are spatially distributed as a homogeneous Poisson point process (HPPP) with intensity $\lambda$. Users are assumed to be spatially distributed as an independent stationary point process with density $\lambda_u$. All BSs are assumed to have users to serve and to continuously transmit at all time, i.e., $\lambda_u \gg \lambda$. The large scale channel gain is captured by the function $L:\mathbb{R}_{+}\rightarrow \mathbb{R}_{+}$ which is assumed to be physically feasible \cite{A_AlAmmouri19} and satisfy the conditions in \cite[Assumption 1]{Area_AlAmmouri20}, for example, the average received power must be less than the transmit power which is assumed to be finite. The bounded single-slope, multi-slope \cite{Wireless_Goldsmith05}, and stretched exponential \cite{SINR_AlAmmouri17} path loss models -- in addition to the path loss models used in 3GPP standards \cite{3GPP2017} for the entire range of $0.5$ to $100$ GHz bands -- are all included in this class of models. The popular but flawed power-law path loss does not belong to this class, however, due to the singularity at the origin. For technical details on this class of path loss models, please refer to  \cite{A_AlAmmouri19}. All BSs are assumed to be equipped with $N(\lambda)=\zeta\lambda$ antennas, where $\zeta>0$, while the users are equipped with a single omnidirectional antenna. We focus on the case where each BS serves one user, extending the model for the multi-user MIMO case is postponed for future work since it requires a fundamentally different approach. 

All small scale channel variables are assumed to be independent of the node locations. Recall that our focus is on mmWave channels, and hence, an i.i.d. complex Gaussian channel assumption between the different antennas is inappropriate. Instead, due to the spatial sparsity of the mmWave channel, it is common to abstract the actual array beam pattern for each beam by a step function with a constant main-lobe over the beamwidth and a constant side-lobe otherwise \cite{Modeling_Andrews16,Performance_Thornburg16,Ergodic_Thornburg18}. Hence, the irregularities in the radiation patterns are ignored and the radiation pattern is abstracted by three parameters: the main-lobe gain $G_{\rm max}(N)$, the side-lobe gain $G_{\rm min}(N)$, and the beamwidth $B(N)$. All of these parameters are functions of the number of antennas and the specific design of the antenna arrays. Typically, the main-lobe gain is non-decreasing with $N$ while the side-lobe gain and the beamwidth are non-increasing \cite{Modeling_Andrews16}. However, deriving the scaling laws with just these monotonicity assumptions is not feasible. To this end, we adopt the following assumptions which we  verify later.

% \begin{figure}[t]
% 		\centerline{\includegraphics[width= 2.8in]{./Figs/ABF2.eps}}
% 		\caption{\, The simplified antenna radiation pattern. The main-lobe gain is $4$dB, the side-lobe gain is $-3$dB, and the beamwidth is $40^{\circ}$.}
% 		\label{fig:mmWave_beam}
% \end{figure}

The main-lobe gain for the antenna array is assumed to be non-decreasing in the number of antennas with $\lim\limits_{N \rightarrow \infty}G_{\rm max}(N)=\infty$. Furthermore, the ratio between the main-lobe gain and the side-lobe gain is assumed to be non-decreasing and to scale with the number of antennas as $\frac{G_{\rm max}(N)}{G_{\rm min}(N)}=\alpha N$, where $\alpha>0$. Finally, the beamwidth is assumed to linearly decrease with the number of antennas, i.e.,  $B(N)=\frac{\beta}{N}$, where $\beta>0$. Based on these assumptions and by conditioning on the network geometry and the channel gains, the conditional SINR given a serving distance of $r_0$ is\footnote{The serving distance has a probability distribution function (PDF) given by $f_R(r_0)= 2 \pi \lambda r_0 e^{- \pi \lambda r_0^2}$ \cite{A_Andrews11}.}
\begin{equation}\label{Eq:SINR_mmWave}
    {\rm SINR}(\lambda)=\frac{ L(r_0)G_{\rm max}(\lambda)\tilde{h}}{\tilde{I}(\lambda)+\bar{I}(\lambda)+\sigma^2 },
\end{equation}
where $\tilde{I}(\lambda)=\sum_{r_i\in \tilde{\Phi}}G_{\rm max}(\lambda)L(r_i)h_i$ is the interference received from BSs with beams pointing towards the tagged user, $\bar{I}(\lambda)=\sum_{r_i\in \bar{\Phi}}G_{\rm min}(\lambda)L(r_i)g_i$ is the interference received from BSs with beams pointing away from the tagged user,  and $\bar{\Phi}$ $\left(\tilde{\Phi}\right)$ is the set of interfering BSs such that the tagged user is in the sight of their main-lobe (side-lobe). Meanwhile, $\tilde{h}$, $h_i, \ \forall i \in \left\{1,2, \cdots \right\}$, and $g_i, \ \forall i \in \left\{1,2, \cdots \right\}$, are independent random variables that capture the small scale fading of the desired link, of interfering links with beams pointing towards the tagged user, and of interfering links with beams pointing away from the tagged user, respectively. Note that these random variables do not need to have identical distributions. Typically, the distribution of $\tilde{h}$, and sometimes the distribution of $h_i, \ \forall i \in \left\{1,2, \cdots \right\}$, is chosen to reflect the LoS nature of the desired link, for example a Ricean or Nakagami distribution, 
while the NLOS $g_i \ \forall i \in \left\{1,2, \cdots \right\}$, are assumed to be Rayleigh fading channels \cite{Modeling_Andrews16}. However, we do not make any assumptions regarding the distributions, except that they should be independent with unit means and independent of the nodes locations. Note that the average SINR can be found by averaging \eqref{Eq:SINR_mmWave} over all channel realizations and network configurations. 

Using this definition of the SINR, the per-user throughput is defined as $\mathbb{E} \left[\log_2(1+ {\rm SINR})\right]$ in bps/Hz and the ASE, which is the network sum throughput per unit area, is given by
\begin{align}\label{Eq:ASEGen}
   \mathbb{E} \left[\mathcal{E}(\lambda) \right]  =\mathbb{E}\left[\lambda \log_2(1+ {\rm SINR}(\lambda)) \right],
\end{align}
in bps/Hz/m$^{2}$.
\section{Performance Analysis}\label{Sec:PerAnalysis}
Using the SINR expression in \eqref{Eq:SINR_mmWave}, we can derive the SINR scaling laws. 
\begin{theorem}\label{Thm:SINRmmWave}
The conditional SINR as defined in \eqref{Eq:SINR_mmWave}, is finite and independent of $\lambda$, as $\lambda \rightarrow \infty$.  Specifically,
     \begin{align}\label{Eq:SINRScalingmmWave}
      \lim\limits_{\lambda \rightarrow \infty}{\rm SINR}(\lambda) &=\frac{L_0 \tilde{h}}{ \sum\limits_{r_i\in \Psi}  L(r_i)h_i+\frac{2 \pi \gamma}{\alpha\zeta} },
      \end{align}
      where $\Psi$ is a PPP with density $\frac{\beta}{2 \pi \zeta}$ and $\gamma=\int\limits_{0}^{\infty} r L(r)dr$, which is finite for physically feasible path loss functions. The mean SINR satisfies:
      \begin{align}\label{Eq:avgSINRScalingmmWave}
          &\lim\limits_{\lambda \rightarrow \infty}\mathbb{E}[{\rm SINR}(\lambda)]=\notag\\
          &\int\limits_{0}^{\infty}L_0 \exp \left( \frac{-2 \pi \gamma t}{\alpha \zeta}-\frac{\beta}{\zeta} \int\limits_{0}^{\infty}\mathbb{E}_h\left[ \left (1-e^{-t h L(r)}\right)\right]r {\rm d}r\right) {\rm d}t ,
      \end{align}
      where the limit is finite and bounded:
      \begin{align}\label{Eq:avgSINRScalingmmWaveBounds}
          \frac{L_0 \alpha \zeta}{\gamma(2 \pi+\alpha \beta) }\leq \lim\limits_{\lambda \rightarrow \infty}\mathbb{E}[{\rm SINR}(\lambda)]\leq \frac{L_0 \alpha \zeta}{2 \pi \gamma}.
      \end{align}
      \begin{proof}
      Refer to Appendix \ref{app:SINRmmWave}.
      \end{proof}
      \end{theorem}
       
Theorem \ref{Thm:SINRmmWave} shows that the linear scaling of the number of antennas is also sufficient to prevent the SINR from dropping to zero in mmWave cellular networks. Note that the theorem focuses on the case where the number of antennas scales linearly with $\lambda$ and it is specific to the assumptions we made regarding the beamwidth $B(N)=\frac{\beta}{N}$ and the antenna gains $\frac{G_{\rm max}(N)}{G_{\rm min}(N)}=\alpha N$. 

The first term in the denominator of \eqref{Eq:SINRScalingmmWave} is due to the interference from BSs with beams pointed towards the tagged user and it is related to the first interference term in \eqref{Eq:SINR_mmWave}. Hence, if the beamwidth $B(N)$ decreased at a rate faster than linear, the first interference term tends zero in the limit, instead of $\sum\limits_{r_i\in \Psi}L(r_i)h_i$. The ratio $\frac{G_{\rm max}(N)}{G_{\rm min}(N)}$ affects the second interference term in \eqref{Eq:SINRScalingmmWave} which reduces to $\frac{2 \pi \gamma}{\alpha\zeta}$ in the limit, when the ratio scales linearly with $\lambda$. If the antenna design allowed the ratio $\frac{G_{\rm max}(N)}{G_{\rm min}(N)}$ to scale at a rate faster than linear, then this interference term would approach zero in the limit. On the other hand, if the ratio scales sub-linearly, then the interference term would grow unbounded in the limit and the average SINR will drop to zero. Based on this, one can have different scaling laws depending on the antenna design itself. Moving to the ASE, the scaling law is given in the next theorem.

\begin{theorem}\label{Thm:ASEmmWave}
The mean ASE, defined in \eqref{Eq:ASEGen}, scales as  $\lim\limits_{\lambda \rightarrow \infty} \mathbb{E}\left[\mathcal{E}(\lambda)\right]=\Theta(\lambda)$. Specifically, we have the following bounds.
\begin{align}
   \frac{1}{\ln(2)}\frac{ \zeta \alpha L_0}{\gamma\alpha\beta +2 \pi \gamma+L_0\alpha \zeta }\leq  \lim_{\lambda \rightarrow \infty}\mathbb{E}\left[\frac{\mathcal{E(\lambda)}}{\lambda}\right]\leq \frac{L_0 \alpha \zeta}{2 \pi \gamma\ln(2)},\notag
\end{align}
where both of these bounds are non-zero and finite constants.
\begin{proof}
Refer to Appendix \ref{app:ASEmmWave}.
\end{proof}
\end{theorem}
Hence, by linearly scaling the number of antennas, one can maintain the desired linear growth of the ASE. {One can also interpret the previous results from a different perspective. If we fix physically size of the antenna arrays, then to be able to fit the desired number of antennas within the array, the carrier frequency has to scale as $\lambda$ for the linear arrays and as $\sqrt{\lambda}$ for the 2D arrays. This argument motivates using higher frequency bands to avoid the densification plateau.}
\begin{figure*}[t]
\centering
		\begin{subfigure}{.5\textwidth}
				% \centerline{\includegraphics[width= 3.15in]{./Figs/Final/SINR_Analog.eps}}
				\centerline{\includegraphics[width= 3.2in]{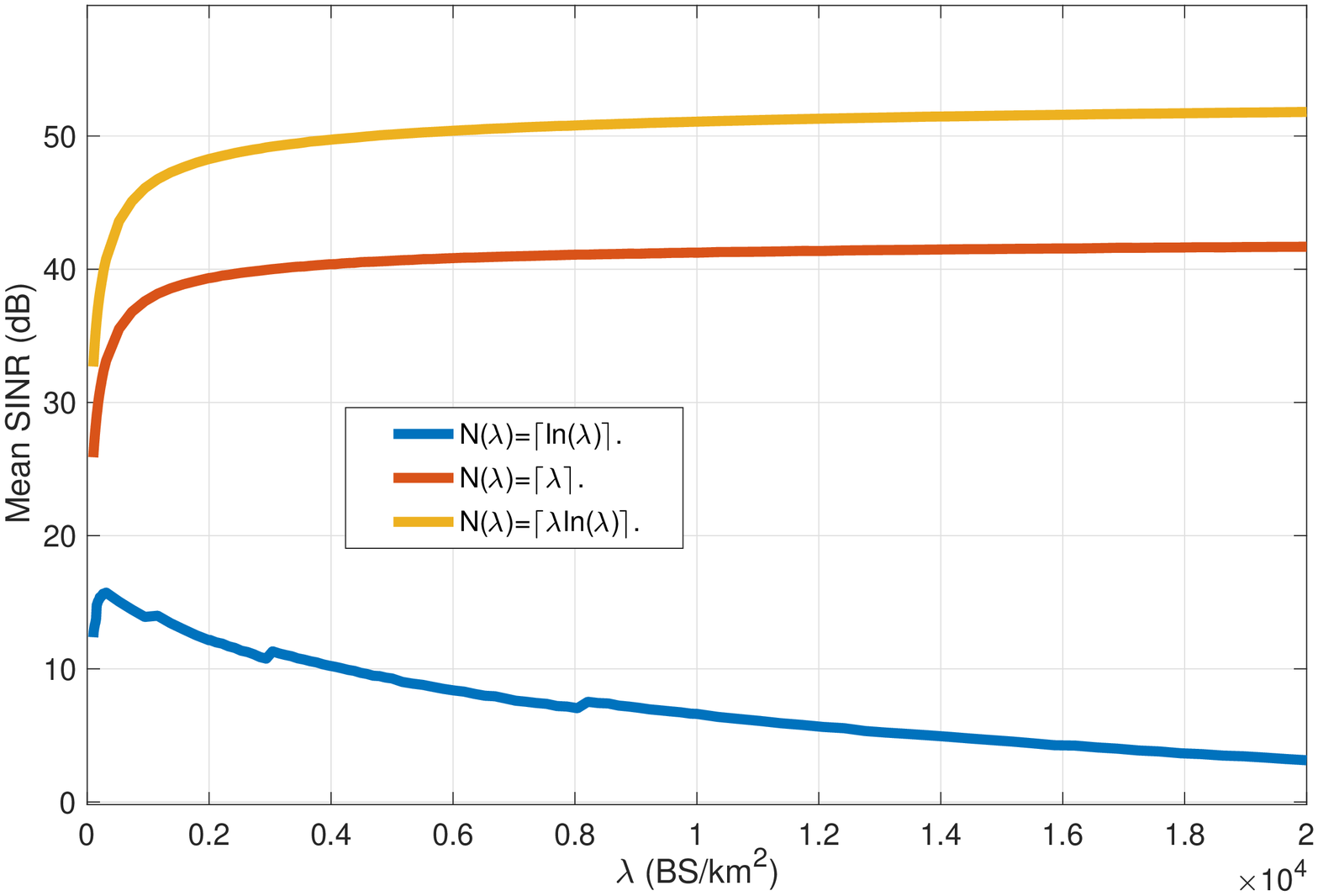}}
		\caption{\, Average SINR.}
		\label{fig:SINR_Analog}
		\end{subfigure}%
		\begin{subfigure}{.5\textwidth}
        % \centerline{\includegraphics[width= 3.4in]{./Figs/Final/ASE_Analog.eps}}
        \centerline{\includegraphics[width= 3.2in]{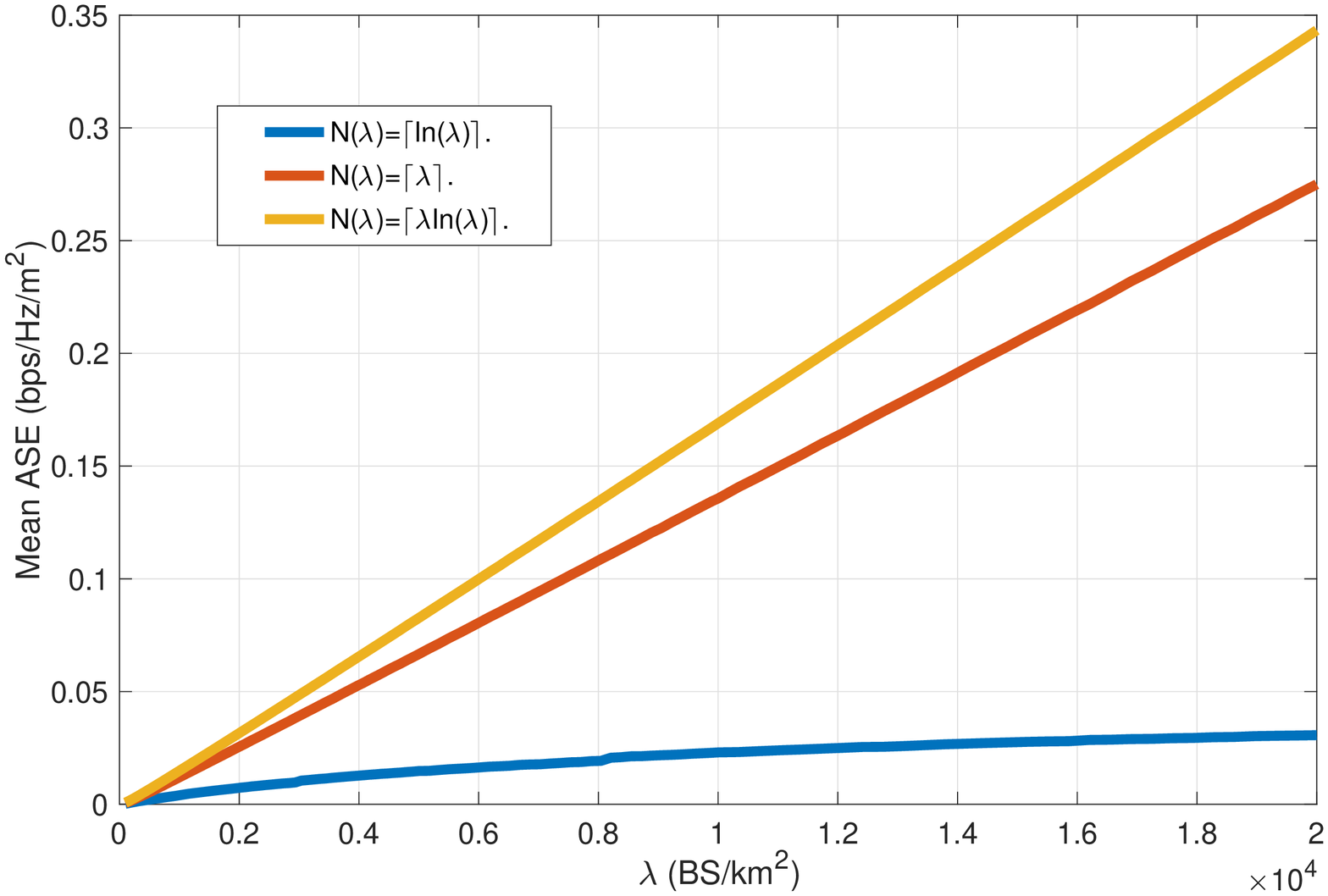}}
		\caption{\, Average ASE.}
		\label{fig:ASE_Analog}
		\end{subfigure}
		\caption{Average ASE and SINR vs the BS density $\lambda$ for different scaling of the number of antennas ($N(\lambda)$).}
		\label{fig:Analog}
\end{figure*} 

\section{Case Study}\label{Sec:CaseStudy}
As we mentioned, the scaling laws in the previous section were derived under specific assumptions regarding the scaling of the main-lobe gain, the side-lobe gain, and the beamwidth. In this section, we consider a specific example and illustrate that the aforementioned assumptions are practically feasible.
\subsection{Uniform Linear Arrays}
% \red{Manan, put a short summary here with the equations for the main and side lobes  highlighting that the radiated power is constant (no need for the theorem). One to two paragraphs  should be enough. Also, cite the works that adopted this model.Note that the model in \cite{Ergodic_Thornburg18} does not maintain the constant transmit power.}
We focus on one dimensional uniform linear array (ULA) of antennas. An $N$ element ULA with inter-antenna spacing $d$ is described by the array response vector $\mathbf{a}(\theta)$ representing the array's phase profile as a function of the angular direction of arriving/departing plane wave.
\begin{align}
\label{Eq:ULA_array_response}
    \mathbf{a}(\theta) &= \frac{1}{\sqrt{N}}\left[1, e^{-j2\pi\theta}, e^{-j4\pi\theta},\cdots,e^{-j2\pi(N-1)\theta}\right]^T,
\end{align}
where $\theta$ is the normalized spatial angle of arrival/departure and is related to the physical angle $\phi$ as $\theta = \frac{d}{\lambda_c}\sin\phi$ and $\lambda_c$ is the wavelength of operation. Typically the inter-antenna spacing is chosen to be half of the operational wavelength, $d=\lambda_c/2$. The maximum antenna gain offered by a ULA is $N$ \cite[Section 2.4]{Opt_Array_VanTrees} and the half-power beamwidth  decreases linearly with $N$ \cite[Table 2.2]{Opt_Array_VanTrees}. Specifically, % Thus, for the step function approximation of the beam pattern described in Fig. \ref{fig:mmWave_beam}, the main-lobe gain $G_{\rm max}(N) = N$ and the beamwidth $B(N) = \Theta_h$.
\begin{align}
    G_{\rm max}(N) &= N\\
    B(N) &=  \frac{1.782}{N}.
    % G_{\rm min} &= \frac{P_{\rm tot} - G_{\rm max}\Theta_h}{\pi - \Theta_h}.
\end{align}

Given $G_{\rm max}(N)$ and $B(N)$, $G_{\rm min}(N)$ is then evaluated such that the radiated power in all directions is constant regardless of $N$.  Similar approximations to the antenna beam pattern have been made in \cite{Alkh17,Renzo15}. With this in mind, the side-lobe gain is
% \begin{align}
% \label{Eq:Side-lobe-EQ}
%     G_{\rm max}(N)B(N) + G_{\rm min}(N)(\pi - B(N)) &= \int_{-\pi/2}^{\pi/2}G_N(\theta)\cos(\theta){\rm d}\theta.
% \end{align}
% where $G_N(\theta)$ is the antenna gain in direction $\theta$. For a ULA the right hand of \eqref{Eq:Side-lobe-EQ} with the array response vector given by \eqref{Eq:ULA_array_response} evaluates to $2$. Thus the side-lobe gain is given by
\begin{align}
    G_{\rm min} = \frac{0.218N}{\pi N - 1.782}.
\end{align}

Hence, all the assumptions we made in the previous section are satisfied asymptotically for the case of ULA.
\subsection{Simulation Results}
Now we illustrate the scaling laws we derived. The simulation drops BSs according to the desired density in a 20x20 km$^2$ region, then for each BS the antenna gain is $G_{\rm max}$ w.p $\frac{B(\lambda)}{2\pi}$ and $G_{\rm min}$ otherwise, where $G_{\rm min}$, $G_{\rm max}$, and $B$ are selected for the ULA discussed before. For the path loss, we use the model adopted in the 3GPP standard \cite[page 26]{3GPP2017} for urban environment (UMa) which incorporates the effects of blockage, the multi-slope nature of the path loss, and the elevation difference between the BSs and the users. This path loss model in included class of physically feasible path loss model we consider as shown in \cite{A_AlAmmouri19}.

The simulation results are shown in Fig. \ref{fig:SINR_Analog} and Fig. \ref{fig:ASE_Analog} assuming three different scaling laws of the number of antennas; sub-linear, linear, and super-linear. Note that in the analysis, we focused on the linear case, so we added the other two in the figures for completeness. The curves agree with our derivation and show that a linear scaling of the number of antennas is required to maintain a non-zero SINR and a linear scaling of the ASE. {Note that in the case of sub-linear scaling, the SINR keeps decreasing for high BSs density and eventually hits zero while the ASE saturates to a constant.} Overall, our results show that the scaling laws observed for the mmWave case with the idealized antenna gain pattern are the same as the ones we derived in \cite{Area_AlAmmouri20} for the traditional cellular networks. The relaxation of the typical i.i.d. assumption for the channel gains adds practical value to the obtained results, especially that the same scaling laws hold in the mmWave bands as well.

\section{Conclusion}\label{Sec:Conc}
In this letter, we studied the scaling laws of the ASE and the SINR in mmWave cellular networks w.r.t. the BS spatial density ($\lambda$), where the BSs are equipped with a number of antennas that scales linearly with $\lambda$. Under fairly general assumptions on the network and signal propagation models, we proved in the limit of dense network, the SINR is non-zero and finite, while the ASE scales linearly with $\lambda$.

\appendices

\section{Proof of Theorem~\ref{Thm:SINRmmWave}}\label{app:SINRmmWave}

Define the following,
\begin{align}
 I(\lambda)&:=\sum_{r_i\in \Phi}G_{\rm min}(\lambda)L(r_i)g_i\notag\\
    I_1(\lambda)&:=\sum_{r_i\in \tilde{\Phi}}G_{\rm max}(\lambda)L(r_i)h_i\notag\\
     I_2(\lambda)&:=\sum_{r_i\in \tilde{\Phi}}G_{\rm min}(\lambda)L(r_i)g_i,\notag
\end{align}
 where $\Phi=\tilde{\Phi}+\bar{\Phi}$. Using these definitions, the SINR in \eqref{Eq:SINR_mmWave} can be written as
 \begin{equation}\label{Eq:AnaSINR2}
    {\rm SINR}(\lambda)=\frac{ L(r_0) G_{\rm max}(\lambda)\tilde{h}}{I(\lambda)+I_1(\lambda)-I_2(\lambda)+\sigma^2 },\notag
\end{equation}
 where $\Phi=\tilde{\Phi}+\bar{\Phi}$. Under the assumption that each BS randomly and uniformly points its beam spatially, one can exploit the independent thinning property of PPPs \cite{Stochastic_Baccelli10_2} and deduce that the density of $\tilde{\Phi}$ is $\lambda\frac{B(\lambda)}{2 \pi}=\frac{\beta}{2 \pi \zeta}$, where the equality holds since we assumed that $B(\lambda)=\frac{\beta}{\lambda \zeta}$. Based on this, the limit of the SINR can be written as
\begin{align}
    \lim\limits_{\lambda \rightarrow \infty} {\rm SINR}(\lambda)&=\frac{ L_0 \tilde{h}}{\lim\limits_{\lambda \rightarrow \infty}\left( \frac{I(\lambda)}{G_{\rm max}(\lambda)}+\frac{I_1(\lambda)}{G_{\rm max}(\lambda)}-\frac{I_2(\lambda)}{G_{\rm max}(\lambda)}\right) }.\notag
    \end{align}
    
    First, note that $\frac{I_1(\lambda)}{G_{\rm max}(\lambda)}=\sum_{r_i\in \tilde{\Phi}}L(r_i)h_i$ which is independent of $\lambda$ and almost surely (a.s.) finite according to the third property of the physically feasible path loss models \cite{A_AlAmmouri19}. Note that $\tilde{\Phi}$ is independent of $\lambda$ and has a density $\frac{\beta}{2 \pi \zeta}$ which is finite. Hence, since $ \lim\limits_{\lambda \rightarrow \infty} \frac{G_{\rm max}(\lambda)}{G_{\rm min}(\lambda)}=\infty$ and $\sum_{r_i\in \tilde{\Phi}} L(r_i)g_i$ is finite a.s., then
 \begin{align}
    \lim\limits_{\lambda \rightarrow \infty} \frac{I_2(\lambda)}{G_{\rm max}(\lambda)}&=\lim\limits_{\lambda \rightarrow \infty}\frac{G_{\rm min}(\lambda)}{G_{\rm max}(\lambda)} \sum\limits_{r_i\in \tilde{\Phi}} L(r_i)g_i=0,\notag
 \end{align}       
For $I(\lambda)$, we have the following 
 \begin{align}
      \lim\limits_{\lambda \rightarrow \infty} \frac{I(\lambda)}{G_{\rm max}(\lambda)}&=\lim\limits_{\lambda \rightarrow \infty}\sum\limits_{r_i\in \Phi}\frac{G_{\rm min}(\lambda)}{G_{\rm max}(\lambda)}L(r_i)g_i\notag\\ 
      &=\lim\limits_{\lambda \rightarrow \infty}\frac{1}{\alpha \zeta \lambda} \sum\limits_{r_i\in \Phi}L(r_i)g_i=\frac{2 \pi \gamma}{\alpha\zeta}\notag,
 \end{align}
 where the last step follows the superposition property of PPP, the law of large numbers, and then by Campbell's theorem as in \cite[Lemma 1]{Area_AlAmmouri20}. Hence,
     \begin{align}
      \lim\limits_{\lambda \rightarrow \infty}{\rm SINR}(\lambda) &=\frac{L_0\tilde{h}}{ \sum\limits_{r_i\in \bar{\Phi}}L(r_i)h_i+\frac{2 \pi \gamma}{\alpha\zeta} },\notag
       \end{align}
which is finite a.s. and independent of $\lambda$. This concludes the asymptotic analysis of the conditional SINR. To show the asymptotic scaling for the mean SINR, we need to show that  $ \lim_{\lambda \rightarrow \infty}\mathbb{E}\left[ {\rm SINR}(\lambda)\right]=  \mathbb{E}\left[\lim_{\lambda \rightarrow \infty} {\rm SINR}(\lambda)\right]$ which follows from the dominated convergence theorem as follows.
\begin{align}
 {\rm SINR}(\lambda)\leq \frac{ L(r_0)G_{\rm max}(\lambda)\tilde{h}}{\tilde{I}(\lambda)+\sigma^2 }\leq  \frac{ L_0 \tilde{h}}{\sum\limits_{r_i\in \tilde{\Phi}}L(r_i)h_i},\notag
\end{align}
where the first inequality holds since $I_2(\lambda)\leq I(\lambda)$ and the second by neglecting the noise term. To apply the dominated convergence theorem, we need to show that the right-hand-side (RHS) term has a finite mean, which follows as
\begin{align}
   \mathbb{E}\left[ \frac{ L_0 \tilde{h}}{\sum\limits_{r_i\in \tilde{\Phi}}L(r_i)h_i}\right]&= \mathbb{E}\left[  \frac{ L_0 }{\sum\limits_{r_i\in \tilde{\Phi}}L(r_i)h_i}\right],\label{Eq:firstMoment}
\end{align}
where the equality holds since $h_i$ has a unit mean. Since $\tilde{\Phi}$ has a finite density, the RHS in \eqref{Eq:firstMoment} is finite, since it was shown in \cite{A_AlAmmouri19} that the RHS  has a finite second moment. Hence, we can conclude that
\begin{align}
      \lim\limits_{\lambda \rightarrow \infty} \mathbb{E} \left[ {\rm SINR}(\lambda)\right] &=\mathbb{E} \left[ \frac{L_0 \tilde{h}}{ \sum\limits_{r_i\in \Psi} h_i L(r_i)+\frac{2 \pi \gamma}{\alpha\zeta} }\right],\notag
       \end{align}
       where $\Psi$ is a PPP with density $\frac{\beta}{2 \pi \zeta}$.  The lower bound in \eqref{Eq:avgSINRScalingmmWaveBounds} is found using Jensen's inequality then Campbell's theorem \cite{Stochastic_Baccelli10_2}, and the upper bound by neglecting the term $\sum\limits_{r_i\in \Psi} h_i L(r_i)$. The exact value is found using the probability generating functional of a PPP \cite{Stochastic_Baccelli10_2}.
       
       \section{Proof of Theorem~\ref{Thm:ASEmmWave}}\label{app:ASEmmWave}
       For this proof, we need to show that, asymptotically, the ASE scales linearly with the BS density, i.e.,  $\lim_{\lambda \rightarrow \infty}\mathbb{E}\left[\frac{\mathcal{E(\lambda)}}{\lambda}\right]=c\in \mathbb{R}_{+}$. To this end, we utilize
       the following bounds.
       \begin{align}
           \lim\limits_{\lambda \rightarrow \infty}\mathbb{E}\left[\log_2(1+{\rm SINR(\lambda)})\right]&\leq \lim\limits_{\lambda \rightarrow \infty} \frac{ \mathbb{E}\left[{\rm SINR(\lambda)}\right] }{\ln(2)}\notag \\
           &\leq \frac{L_0 \alpha \zeta}{2 \pi \gamma\ln(2)},\notag
       \end{align}
       where the first upper bound holds since  $\log_2(1+x)\leq\frac{x}{\ln(2)}$ and second bound follows from \eqref{Eq:avgSINRScalingmmWaveBounds}. We can also use Fatou's lemma to obtain the following lower bound.
       \begin{align}
           \lim\limits_{\lambda \rightarrow \infty}&\mathbb{E}\left[\log_2(1+{\rm SINR(\lambda)})\right]\geq  \mathbb{E}\left[\lim\limits_{\lambda \rightarrow \infty}\log_2(1+{\rm SINR(\lambda)})\right]\notag\\
           &=\mathbb{E}\left[\log_2\left(1+\frac{L_0 \tilde{h}}{ \sum\limits_{r_i\in \Psi}  L(r_i)h_i+\frac{2 \pi \gamma}{\alpha\zeta} }\right)\right] \notag\\
           &\geq \frac{1}{\ln(2)}\mathbb{E}\left[\frac{L_0}{ \sum\limits_{r_i\in \Psi}  L(r_i)h_i+\frac{2 \pi \gamma}{\alpha\zeta}+L_0 \tilde{h}}\right] \label{eq:ASEPro1}\\
           &\geq \frac{1}{\ln(2)} \frac{L_0}{\mathbb{E}\left[ \sum\limits_{r_i\in \Psi}  L(r_i)h_i+\frac{2 \pi \gamma}{\alpha\zeta}+L_0 \tilde{h}\right]}\label{eq:ASEPro2} \\
           &= \frac{1}{\ln(2)}\frac{L_0}{\frac{\gamma\beta}{\zeta} +\frac{2 \pi \gamma}{\alpha\zeta}+L_0 }, \label{eq:ASEPro3}
       \end{align}
       where \eqref{eq:ASEPro1} follows since $\ln(1+x)\geq \frac{x}{1+x}$, i.e., $\log_2(1+x)\geq \frac{x}{(1+x)\ln(2)} $, \eqref{eq:ASEPro2} follows from Jensen's inequality, and \eqref{eq:ASEPro3} follows from Campbell's theorem. Note that \eqref{eq:ASEPro3} is a non-zero finite constant and independent of $\lambda$. Hence, $\lim_{\lambda \rightarrow \infty}\frac{\mathbb{E}\left[\mathcal{E(\lambda)}\right]}{\lambda}$ is bounded by non-zero finite constants from below and above, which concludes the proof. 
     
\bibliographystyle{IEEEtran}
\bibliography{AhmadRef}
\vfill
\end{document}